\documentclass[aps,twocolumn,showpacs]{revtex4}

\usepackage{epsfig}
\usepackage{epsf}

\def\slash#1{\setbox0=\hbox{$#1$}
   \dimen0=\wd0 \setbox1=\hbox{/} \dimen1=\wd1
   \ifdim\dimen0>\dimen1 \rlap{\hbox to \dimen0{\hfil/\hfil}} #1
   \else  \rlap{\hbox to \dimen1{\hfil$#1$\hfil}} / \fi}
   
\begin{document}

\title{On the coupling constant bounds implying area decay of Wilson loop for Z2 lattice gauge theories with Wilson action Boltzmann factor given by bond variables described by transverse Ising model.}

\author{F. C. S\'{a} Barreto}
\email{fcsabarreto@gmail.com}
\author{A. L. Mota}
\email{motaal@ufsj.edu.br}
\affiliation{Departamento de Ci\^{e}ncias Naturais, Universidade Federal de S\~{a}o Jo\~{a}o del Rei,
C.P. 110,  CEP 36301-160, S\~ao Jo\~ao del Rei, Brazil}

\begin{abstract}
We obtain correlation functions for lattice gauge theories with action Boltzmann factor given by bond variables described by transverse Ising model
and apply them to obtain area decay of the Wilson loop observable in a range of the coupling parameter larger than that obtained from mean field theory considerations.
\end{abstract}

%64.60.De Statistical mechanics of model systems (Ising model, Potts model, field-theory models, Monte Carlo techniques, etc.)  
%11.15.Ha Lattice gauge theory (see also 12.38.Gc Lattice QCD calculations) 
%75.10.Jm Quantized spin models, including quantum spin frustration  
%05.30.Rt Quantum phase transitions  
%05.50.+q Lattice theory and statistics (Ising, Potts, etc.)  

\maketitle

%\section{Introduction}
In this paper we study the lattice gauge theories using correlation identities and rigorous inequalities to obtain coupling parameters which imply area decay in lattices decorated by bond variables of  spins $\sigma=1/2$ described by the transverse Ising action. Therefore we study the effect of a disordered field on the area decay behavior of the Wilson loop observable. Previously \cite{Nikolic2005}, it has been presented the study of the quantum transverse Ising antiferromagnet on the kagome lattice, with weak transverse field dynamics and
other local perturbations. The possibility of a disordered zero-temperature phase is investigated by means of an appropriate
mapping to a compact U(1) gauge theory on the honeycomb lattice that is coupled to a charge-1 matter field.
The interest on such systems lies on the possibility of exotic phases and spin liquid states appearing in geometrically frustrated quantum magnets, as, for example, in 
\cite{Anderson1987,Senthil2000,Coldea2001,Serbyn2011,Dang2011}. Spin ice \cite{Hertog2000,Bramwell2001}, its mapping on a Ising system on a kagom\'{e} lattice \cite{Tanaka2006,Mellado2010} and excitations on the spin ice ground state \cite{Castelnovo2008} also constitute recent examples. In all these cases, in order to identify phase transitions on the system, the absence of a non-zero magnetization to be employed as an order parameter can be circumvented by evaluating the decay law of the Wilson loop \cite{Tsvelik}.

Here, we consider the well known Wilson loop observable of a pure gauge $Z^d$ lattice theory with Wilson action defined as
\begin{equation}
<W( C )> = \lim_{\Lambda \to Z^d} < W( C) >_\Lambda \label{eq1}
\end{equation}   
where $< W( C) >_\Lambda $, is the finite lattice Gibbs ensemble average, with action Boltzmann factor \cite{Wilson1974,Seiler1982,Kogut1979}, given by,
\begin{equation}
\exp\Big({\beta\Omega\sum_i \sigma _i^x + \beta J\sum_{P\subset\Lambda}{\chi}_P}\Big), \label{eq2}
\end{equation}   
where P denotes the unit squares (plaquettes) of $\Lambda$, and  $\chi _P$ is given by, $\chi _P = \sigma _1^z\sigma _2^z\sigma _3^z\sigma _4^z$; $\sigma _i^z $ and $\sigma_i ^x$ are the Pauli matrix for the spin 1/2. 

The action is, 
\begin{equation}
H = -\Omega\sum_i \sigma _i^x +  \frac{J}{2}\sum_{P\subset\Lambda}\chi _P \label{eq3}
\end{equation}   
 
Let W(C) be the product of $\sigma _b^z$, where $b$ stands for the bond, along the perimeter of the planar rectangle C of area A. The gauge coupling constant is $ J$ , $0 < \beta J< \infty$. $\Omega$ is the  local transverse field in the bond that exchanges its value from  $\pm 1$ to $\mp 1$ .Therefore the sign of the plaquette changes according to the $\Omega  \sigma _i^x$ term of the action Eq.(\ref{eq3}).  Area decay of $<W( C )>$ is a criterion for confinement. 
In section 3, we deduce the identity for $<\sigma _D^z>$, where $\sigma_D^z = \sigma _{i1}^z...\sigma _{iD}^z$ denotes a product of distinct bond variables. Similar procedure has been applied previously for the Z2 lattice gauge Ising case \cite{SaBarreto1983a}. We take free boundary conditions and note that Griffith's first and second inequalities for the spin model apply \cite{Gallavotti1971,Suzuki1973,Contucci2010} and therefore imply the existence of the thermodynamic limit \cite{Glimm1981}. For d = 3 and d = 4 we obtain lower bounds ${\beta_L  }$ on the area decay of $<W( C )>$ ( i.e. for all $\beta <{\beta _L  }$ , $<W( C )>$ has area decay, using correlation equalities for $\sigma _D^z$ and Griffith's inequalities. For completeness, in section 2, we give the mean field lower bound $\beta _M
(\beta _M<\beta _L )$ using a decoupling and Griffith's inequality argument \cite{SaBarreto1983a, Tomboulis1981}.

%\section{ Mean field lower bounds}

For definiteness assume C lies in the $x_1,x_2$  plane. Consider a bond b fixed in the lower left-hand corner of C. Replace $\beta J$ by $(\beta  J)\lambda $, $\lambda  \in  [0,1]$, in the action for the $2(d-1)$ plaquettes (call them $P_1...P_{2(d-1)})$ that have one bond in common with b. In the following we will discard the mechanism of disorder represented by $\Omega$. Denote the corresponding expectation by $<W(C)>_\lambda $ . Integrating $(\frac{d}{d\lambda })<W(C)>_\lambda $ gives, noting that $<W(C)>_0 = 0$,
\begin{eqnarray}
&& <W(C)> =\int_{0}^{1} d\lambda \frac{d}{d\lambda } <W(C)>_\lambda \nonumber \\
&&=\beta \int_{0}^{1} d\lambda \sum_{i=1}^{2(d-1)} (<W(C) \chi _{P_i}>_\lambda  \nonumber\\
&& - <W(C)>_\lambda> < \chi _{P_i}>_\lambda)\nonumber \\
&&\leq \beta \int_{0}^{1} d\lambda \sum_{i=1}^{2(d-1)} <W(C) \chi _{P_i}>_\lambda\nonumber \\
&&\leq \beta \sum_{i=1}^{2(d-1)} <W(C) \chi _{P_i}>,\label{eq4}            
\end{eqnarray} 
using Griffith's� first (second) inequality in the first (second) equality. Each term on the right corresponds to a modified contour determined by the bonds of the variables of $W( C ) \chi _{P_i}$ which enlarges or diminishes C by one plaquette. We repeat the argument proceeding along successive rows of plaquettes enclosed by C. After A applications we arrive at
 \begin{equation}
<W( C )> \leq  \beta^A \text{(sum of } [2(d-1)]^A \text{ terms)} \label{eq5}        
\end{equation}   
Each term is non-negative and bounded above by 1 giving $\Big<W( C )\Big>\leq  [\beta  2(d-1)]^A$.
Therefore, for each $\beta$ , 
\begin{equation}
\beta \in 1/(2(d-1)),\label{eq6} 
\end{equation}
we have, 
\begin{equation}
\Big<W( C )\Big> \leq  e^{ [-\log (2\beta  (d-1)A)]}\label{eq7}.
 \end{equation}
In the preceding arguments, which led to the mean field lower bound $\beta_M$, we have modified the action by replacing $\beta J$ by $(\beta  J)\lambda $. The effect of $\Omega$ is to lower the critical temperature $T_c$, or to increase $\beta_c$, up to $T_c=0$, or $\beta_c=\infty$.   
  
%\section{Calculation of the Identity for $<\sigma _D^z>$ the bond variables product average.}

We now present the derivation of the longitudinal spin correlation function which is derived in a manner analogous to the ones for the spin � transverse Ising systems \cite{SaBarreto2011}(see also \cite{SaBarreto1981}) and they will be used to extend the $\beta $ region of area decay given by relation (\ref{eq6}),  for d = 3 and 4. 
Let $\sigma _D^z = \sigma _{i1}^z...\sigma _{iD}^z$ denote a product of distinct bond variables and for a fixed bond b occurring in $\sigma _D^z$ give a numerical ordering 1,2,... to the 2(d- 1) plaquettes that have one bond in common with b. Let $\sigma _D^{(b)}$ be the product $\sigma _D$ with the bond b deleted.
Let us consider the bond b, with $\sigma _b$, and the plaquette $\chi _b$ which contains the bond b.
Summing over $\sigma _b$, we have,
\begin{equation}
<\sigma _D^z> = \Big<\sigma _D^{{z}{(b)}}\frac{\sum_{j,k,l} J_{ij}\sigma _j^z \sigma _k^z\sigma _l^z}{E_i} \tanh \beta E_i\Big>\label{eq8}
\end{equation}   
where,
\begin{equation}
E_i = \sqrt {E_{i,x}^2 + E_{i,z}^2}\label{eq9}
\end{equation} 
with $E_{i,x} = \Omega$ and $E_{i,z} = \sum_{j,k,l} J_{ij}\sigma _j^z \sigma _k^z\sigma _l^z$. 

Also, $\sigma _j^z $,$\sigma _k^z$ and $\sigma _l^z$ are the neighbours of $\sigma _b^z$ on the plaquette $\chi _b$.  

From the expectation value \begin{equation}<\sigma _D^z> = Tr(\sigma _D^z e^{-\beta H)})/Z\end{equation} we obtain, writing $H = H_b + H'$, where $H_b$ includes all terms related to bond b and $H'$ the rest of the lattice, the equation,
\begin{eqnarray}
&&< \sigma _D^z> = \Big<\sigma _D^{{z}{(b)}} \frac{ Tr_{(b)}\sigma _D^z e^{-\beta H_b}}{Tr_{(b)}e^{-\beta H_b}}\Big> \nonumber \\
&&-\Big<\sigma _D^{{z}{(b)}}\Big[\frac{Tr_{(b)}\sigma _D^z e^{-\beta H_b}}{Tr_{(b)}\sigma _D^z e^{-\beta H_b}}-\sigma _D^z\Big] \Delta \Big>\label{eq10}
\end{eqnarray} 

where $Tr_{(b)}$ represents the partial trace with respect to bond b and $\Delta= 1 - e^{-\beta H_b}e^{-\beta H^\prime}e^{\beta (H_b+H^\prime)}$.
Equation (\ref{eq10}) is an exact relation. However, it is difficult to be used. Therefore, we will make an approximation based on the following decoupling,

\begin{eqnarray}
&&\Big<\sigma _D^{{z}{(b)}}\Big[\frac{Tr_{(b)}\sigma _D^z e^{-\beta H_b}}{Tr_{(b)}e^{-\beta H_b}}-\sigma _D^z\Big]\Delta\Big> \approx \nonumber\\
&&\Big<\sigma _D^{{z}{(b)}}\Big[\frac{Tr_{(b)}\sigma _D^z e^{-\beta H_b}}{Tr_{(b)}e^{-\beta H_b}}-\sigma _D^z\Big]\Big> \Big<\Delta\Big> \label{eq11}
\end{eqnarray} 
 
Inserting $(\ref{eq11})$ into $(\ref{eq10})$ and using the fact that $\Delta \neq 1$, we obtain,

\begin{equation}
<\sigma _D^z> \leq  \Big<\sigma _D^{{z}{(b)}}\frac{ Tr_{(b)} \sigma _D^z e^{-\beta H_b}}{Tr_{(b)} e^{-\beta H_b}}\Big>\label{eq12} 
\end{equation} 
By expanding $\Delta$ we see that the approximation is correct to the order of $\beta^2$ and it is consistent with the application of the correlation inequalities that will be used later. Remarck : A more general result can be represented by equations identical to (\ref{eq10}) and (\ref{eq12}) but written for $<F(\sigma)\sigma_b^z>$, where $F(\sigma)$ is any product function of the bond components, except $\sigma_b^z$. Relations (\ref{eq10}) and (\ref{eq12}) are special cases where $F(\sigma)=\sigma_D^{{z}{(b)}}$. The relation for the longitudinal bond magnetization $<\sigma_b^z>$ is the special case where $F(\sigma)=1$. Within this description, the decoupling (\ref{eq11}) can then be viewed as a $0$-th order approximation of the exact relation (\ref{eq10}) for $F(S)=1$ , i.e. , Eq.(\ref{eq12}) can be assumed to be obtained from the approximation 
$<\bar{\sigma_b^z} - \sigma_b^z>=0$  where $\bar{\sigma_b^z} = Tr_{(b)}\{ \sigma_b^z e^{-\beta H_b} \} / Tr_{(b)}\{ e^{- \beta H_b} \}$.

Introducing $\nabla = \frac{\partial}{\partial x } $ through $e^{\alpha \nabla}.f(x) = f(x+\alpha)$, we get,
\begin{eqnarray}
&&<\sigma _D^z> \leq \Big<\sigma _D^{{z}{(b)}}  e^{(\sum_{j,k,l} J_{b,j,k,l}\sigma _j^z \sigma _k^z\sigma _l^z)\nabla}\Big>f(x)\mid _{{x=0}}\label{eq13}
\end{eqnarray}  
and, 
\begin{equation}
f(x) = \frac{x}{\sqrt{(2\Omega)^2 + x^2}}\tanh(\beta)\sqrt{(2\Omega)^2 + x^2} = -f(-x)\label{eq14}
\end{equation} 
Considering,
\begin{eqnarray}
&& \Big< e^{(\sum_{j,k,l} J_{b,j,k,l}\sigma _j^z \sigma _k^z\sigma _l^z)\nabla}\Big>\nonumber\\
&&=\Big<\prod_{j,k,l}e^{(J_{bjkl}\sigma _j^z \sigma _k^z\sigma _l^z)\nabla}\Big> \nonumber\\
&&= \Big<\prod_{j,k,l}\Big[\cosh({J_{bjkl}\nabla}) + \sigma _j^z \sigma _k^z\sigma _l^z\sinh({J_{bjkl}\nabla)\Big]}\Big> \label{eq15}
\end{eqnarray} 

we obtain,

\begin{eqnarray}
&&<\sigma _D^z> \leq \Big<\sigma _D^{{z}{(b)}} \prod_{j,k,l}\Big[\cosh({J_{bjkl}\nabla})\nonumber\\ 
&&+ \sigma _j^z \sigma _k^z\sigma _l^z\sinh({J_{bjkl}\nabla)\Big]}\Big>.f(x)\mid _{{x=0}} \label{eq16}
\end{eqnarray} 

%\section{ Coupling bounds for the area decay for d=2, d=3 and d=4.}

Applying the previous result, equation (\ref{eq16}), for d= 2,3,4, after some algebra, we obtain, (a), (b) and (c):
\begin{eqnarray}
(a)&&d=2 \nonumber\\
&&<\sigma _D^z> \leq a_2 \sum_i <\sigma _D^z \chi_{P_i}>,\\ \label{eq17}
&&a_2 = \frac{1}{2} f(2 J), a_2\geq 0\nonumber
\end{eqnarray}

\begin{eqnarray}
(b)&&d=3\nonumber\\
&&<\sigma_D^z> \leq a_3 \sum_i <\sigma_D^z \chi_{P_i}>\nonumber\\
&&+ b_3 \sum_{ijk} <\sigma _D^z \chi_{P_i}\chi_{P_j}\chi_{P_k} >,\\\label{eq18}
&&a_3  = \frac{1}{2^3} [f(4 J)+2f(2 J), a_3\geq 0\nonumber\\
&&b_3 = \frac{1}{2^3} [f(4 J)-2f(2 J), b_3\leq  0\nonumber
\end{eqnarray}

\begin{eqnarray}
(c)&&d=4\nonumber\\
&&<\sigma _D^z> \leq a_4 \sum_i <\sigma _D^z \chi_{P_i}>\nonumber\\ 
&&+ b_4 \sum_{ijk} <\sigma _D^z \chi_{P_i} \chi_{P_j}\chi_{P_k} >\nonumber \\
&&+ c_4 \sum_{ijklm} <\sigma _D^z \chi_{P_i}\chi_{P_j}\chi_{P_k} \chi_{P_l}\chi_{P_m} >, \\\label{eq19}
&&a_4  = \frac{1}{2^5} [f(6 J)+4f(4 J)+5f(2 J), a_4\geq 0\nonumber\\
&&b_4 = \frac{1}{2^5} [f(6 J)-3f(2 J), b_4\leq  0\nonumber\\
&&c_4  = \frac{1}{2^5} [f(6 J)-4f(4 J)+5f(2 J), c_4\geq 0\nonumber
\end{eqnarray}

We will now derive the coupling bounds for the area decay of the correlation functions . Applying the procedure described below to equations (\ref{eq18}) and (\ref{eq19}), we will obtain the upper bounds for d=3 and d=4.
If $\beta J$ is such that:
\begin{eqnarray} 
&&(a) 4 a_3 <1,\nonumber \\
&&\text{then } \Big<W( C )\Big> \leq  e^{(-ln[4a_3] A)},   \text{for d= 3}.   \\\label{eq20} 
&&\nonumber\\
&&(b) 6(a_4 + c_4) <1, \nonumber \\
&&\text{then } \Big<W( C )\Big> \leq   e^{(-ln[6(a_4 + c_4] A)},   \text{for d= 4}. \label{eq21}       
\end{eqnarray} 
The proofs of results $(a)$ and $(b)$ follows the same procedure used to obtain the mean field results (section2).
Let us prove result (a). For $b \in C$,
\begin{eqnarray}
&&<W(C)> \leq a_3 \sum_i <W(C)\chi_{P_i}> \nonumber\\
&&+ b_3 \sum_{i<j<k} <W(C) \chi_{P_i}\chi_{P_j}\chi_{P_k} >\nonumber\\
&&\leq a_3 \sum_i <W(C)\chi_{P_i}>\label{eq22}
\end{eqnarray}

The last inequality is obtained because $b_3$ is negative and $<W(C) \chi_{P_i}\chi_{P_j}\chi_{P_k} >$ is positive by Griffith's first inequality. At each application of the equality we pass to an inequality by dropping the $b_3$ terms. After A steps we arrive at
\begin{equation}
<W( C )> \leq  a_3^A \text{(sum of } 4^A \text{ terms)} 
\end{equation}
where each term is less than one.
To prove result (b) for d=4 we use equation $(\ref{eq18})$ and proceed as previously as in the proof of result (a) for d=3 . We can drop the $b_4$ terms in favour of an inequality. At each stage we have six terms from the $a_4$ term and six terms from the $c_4$ term.

The values of $\Omega_c$ , the local disorder critical field, for which the conditions (\ref{eq20}) and (\ref{eq21}) are not satisfied, in other words, for which the perimeter decay regime of the Wilson loop dominates, are:
\begin{eqnarray}
&&\Omega _c (d=3) = 1.3755 J\\
&&\Omega _c (d=4) =2.4466 J
\end{eqnarray}
For values greater than those there are no confinement.

%\section{Concluding Remarks} 

In this paper, we introduce the $Z^d$ lattice gauge model for which the action has a term that represents a disorder mechanism. This term is described by a local field represented by the $\sigma^x$ spin 1/2 operator. The effect of this local field is to change the sign of the plaquette, i.e, from $\pm 1$ to $\mp 1$. We have use correlation identities and  rigorous inequalities, to obtain, for d = 3 and d = 4, lower bounds for the area decay of $<W(C)>$, i.e , for all
$\beta \leq \beta _L$,  $<W(C)>$ has area decay.
The correlation identities are a gauge version of Callen's identities employed by the
authors \cite{SaBarreto2011} to obtain lower than mean field upper bounds for the critical couplings for the transverse Ising model. The procedure was based on an approximation for an exact identity and on rigorous inequalities for the spin correlation functions. We obtain the critical local disorder field above which the perimeter decay regime of the Wilson loop dominates. 

%\section{Acknowledgements}

FCSB is grateful for the financial support of CAPES/Brazil which made possible his visit to the UFSJ/Brasil. ALM acknowledges financial support from CNPq and FAPEMIG (brazilian agencies).

\end{document}